# All-optical scattering control in an all-dielectric quasi-perfect absorbing Huygens' metasurface

*Kentaro Nishida[1], Koki Sasai[2], Rongyang Xu[2], Te-Hsin Yen[1], Yu-Lung Tang[1], Junichi Takahara[2,3,\*], and Shi-Wei Chu[1,4,5,\*]*

[1] Department of Physics, National Taiwan University, No. 1, Sec. 4, Roosevelt Rd., Taipei 10617, Taiwan

[2] Graduate School of Engineering, Osaka University, 2-1 Yamadaoka, Suita, Osaka 565-0871, Japan

[3] Photonics Center, Graduate School of Engineering, Osaka University, 2-1 Yamadaoka, Suita, Osaka 565-0871, Japan

[4] Molecular Imaging Center, National Taiwan University, No. 1, Sec 4, Roosevelt Rd., Taipei 10617, Taiwan

[5] Brain Research Center, National Tsing Hua University, 101, Sec 2, Guangfu Road, Hsinchu 30013, Taiwan

ABSTRACT:

In this paper, we theoretically and experimentally demonstrated photothermal nonlinearities of both forward and backward scattering intensities from quasi-perfect absorbing silicon-based metasurface with only $\lambda/7$ thickness. The metasurface is efficiently heated up by photothermal effect under laser irradiation, which in turn modulates the scattering spectra via thermo-optical effect. Under a few milliwatt continuous-wave excitation at the resonance wavelength of the metasurface, backward scattering cross-section doubles, and forward scattering cross-section reduces to half. Our study opens up the all-optical dynamical control of the scattering directionality, which would be applicable to silicon photonic devices.

1. **Introduction**

Dielectric nanostructures with high refractive index efficiently support Mie resonance, such as electric dipole (ED) and magnetic dipole (MD), in the visible wavelength region, and have enabled nano-scale control of electromagnetic waves by exploiting the strong light confinement effect [1]. In particular, the metasurface, which is constructed by the periodic dielectric nanostructures (meta-atoms), allows us to efficiently tune the wavelengths of Mie resonance modes by adjusting the size and the period of meta-atoms, exploiting collective resonance caused by the electromagnetic couplings between meta-atoms in the metasurface [2–4]. When the metasurface excites the degenerate ED and MD modes (i.e., Huygens' dipole), light reflection from the metasurface diminishes by destructive interference [5,6] and thus the metasurface has strong forward scattering directionality, leading to so-called the Huygens' metasurface [2,7], with various applications including optical antenna and anti-reflection coating [8,9].

    Recently, several techniques for the dynamic tuning of scattering direction in the dielectric Huygens' metasurfaces were proposed. One major approach is to use phase-change material, which enables reversibly tuning the resonance modes by switching between amorphous and crystalline states via heating [10,11]. For example, $Ge_2Se_2Te_5$ nanodisc metasurface demonstrated the capability of red-shifting the resonance wavelength of Huygens' dipole by increasing crystalline fraction [12]. The other approach is to exploit the thermal modulation of the refractive index on dielectric materials to tune the resonance modes. In particular, taking advantage of the large thermo-optical effect of silicon, reversible controls of forward and backward scattering ratios have been achieved by varying the temperature of silicon metasurfaces [13,14]. These thermo-optically active control techniques are significant contributions to the development of

nanophotonic devices. Toward practical applications, all-optical control is a highly desirable milestone [15,16]. We have recently discovered giant photothermal nonlinearity that allows all-optical switching of scattering in a silicon nanoparticle [17,18]. However, all-optical Huygens' metasurface scattering control combining photo-thermal and thermo-optical effects has not been realized.

In this paper, we theoretically and experimentally demonstrated photothermal nonlinear optical effects from a silicon Huygens' metasurface, enabling all-optical controls of both forward and backward scattering intensities. Figure 1 shows the concept. Initially, the forward scattering spectrum has a peak at the resonance wavelength, while the backward scattering spectrum shows a dip, due to the property of the Huygens' metasurface. When the metasurface absorbs energy from incident light, the complex refractive index of silicon is modified due to temperature rise [17,19], leading to spectral red-shift in both forward and backward scattering spectra. With a fixed excitation wavelength at the resonance peak of the metasurface, the forward scattering cross-section decreases, and the backward scattering cross-section increases. That is, the photothermally modulated Huygens' metasurface induces nonlinear scattering intensity dependencies on the excitation intensity in both forward and backward directions. In our research, to improve the efficiency of the photothermal effect, we used quasi-perfect absorbing silicon metasurface, which is designed to approach the condition of degenerate critical couplings [20,21]. Our study opens up the all-optical dynamical control of the scattering directionality, which extends the functionality of silicon photonic devices.

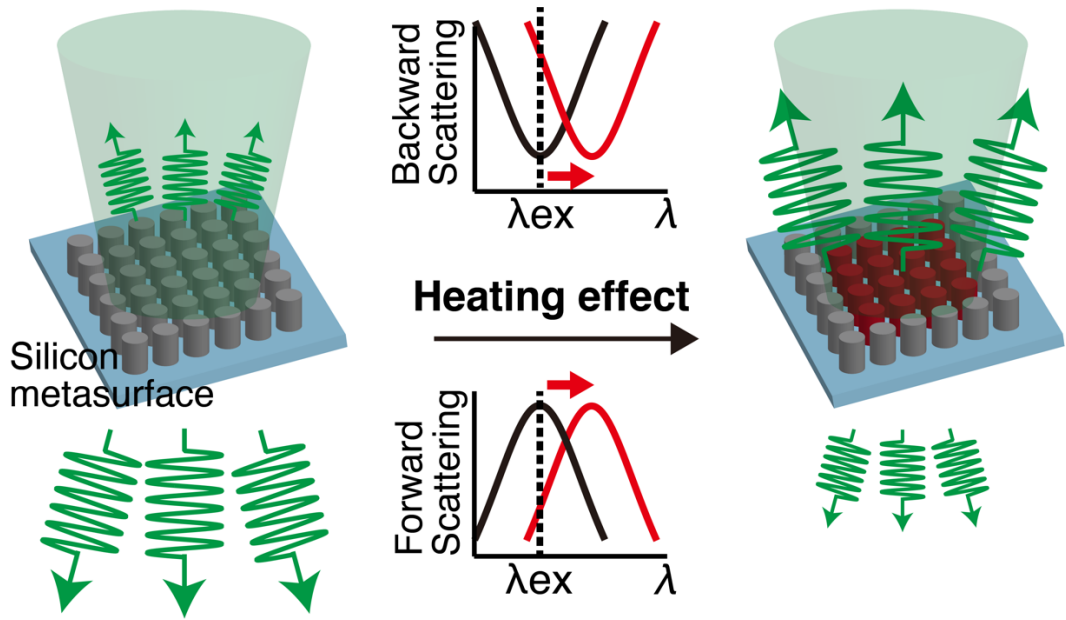

Figure 1: Mechanism on photothermal control of scattering directionality from silicon metasurface. The green cone is the incident beam, and the green arrows represent scattering light. Under resonant excitation, the quasi-perfect absorber metasurface is efficiently heated up by photothermal effect, and subsequent thermo-optical effect results in spectral red-shift, thus changing scattering directionality.

2. Methods and results

**2.1 Fabrication of silicon metasurface**

Figure 2a shows the structure of the silicon metasurface. The meta-atoms of crystalline silicon nanodisc resonators with a diameter of 200 nm and a height of 79 nm are periodically aligned at the period of 305 nm. The silicon metasurface was fabricated on the single crystalline silicon on a quartz substrate. The single crystalline silicon on a quartz substrate was a custom-made product by Shin-Etsu Chemical Co., Ltd, which is fabricated by the bonding process of the silicon and quartz wafers after implementing $H^+$ ion to the surface of the silicon wafer. The thickness of the silicon layer is adjusted to 79 nm by the etching process using SF6 and C4F8 plasma gasses in a reactive ion chamber. The meta-atom structural pattern was drawn by using electron beam lithography (ELS-7700T, Elionix Inc.) on a chemical resist (SEP 520 A, Zeon Corp.) that is spin-coated on the silicon surface. The lift-off procedure was assisted by evaporating a 30 nm thick Cr mask on the electron beam lithography pattern of the chemical resist. The silicon layer that was not covered by the Cr mask was selectively etched by using plasma gasses. The remaining Cr mask on the silicon metasurface was then removed by immersing the substrate in di-ammonium cerium (IV) nitrate solution. We confirmed that the structure of silicon metasurface is properly fabricated via a scanning ion microscope (SIM), as shown in Figure 2b.

Figure 2c shows the simulated absorption spectral map with various heights of silicon meta-atoms on metasurface, calculated by a commercial finite-difference time-domain (FDTD) simulator (Lumerical, Ansys Inc.). When the height of silicon meta-atoms is 79 nm, degenerate coupling occurs (ED and MD simultaneously excited) at the wavelength of 561 nm. In addition, the absorption of the silicon metasurface is maximized by tuning the period of the silicon nanodisc meta-atoms so that the condition of critical coupling, where the radiative energy loss rate and intrinsic energy loss rate are balanced in the resonator [22,23]. In the critical coupling condition, the transmitted light is eliminated due to the destructive interference with the internal field in the resonator, and the resonator exhibits the maximum absorption. Our calculation found the period of 305 nm, which is the same as our design, provides the highest absorption of 90% at the wavelength of 561 nm, as shown in Supplementary Materials (Figure S1). Note that the reason why the theoretical absorption did not reach 100% is that the critical coupling condition was not completely fulfilled due to the insufficient imaginary permittivity of crystalline silicon at 561 nm [20]. Figure 2d shows the nice agreement between the experimentally measured absorption spectrum from the fabricated silicon metasurface and the calculated spectrum. The experimental procedure of the spectral measurement is shown in Figure S2 of Supplementary Materials. Both spectra exhibit sharp absorption at 561 nm, indicating our silicon metasurface is indeed working as a quasi-perfect absorbing metasurface, though the experimental peak absorption is ~10% degraded compared to that of the theoretical spectrum, probably due to minor fabrication inaccuracy. It should be noted that another high absorption region is seen at the wavelength of 450 nm by the complex mixing of other resonance modes, such as quadrupole modes. Because this resonance does not fulfill the condition of Huygens' metasurface, no specific scattering directionality is induced at the 450 nm wavelength. However, it is possible to achieve all-optical scattering modulation by exploiting the fact that high absorption induces strong photo-thermal and thermo-optical effects on the silicon nanostructure.

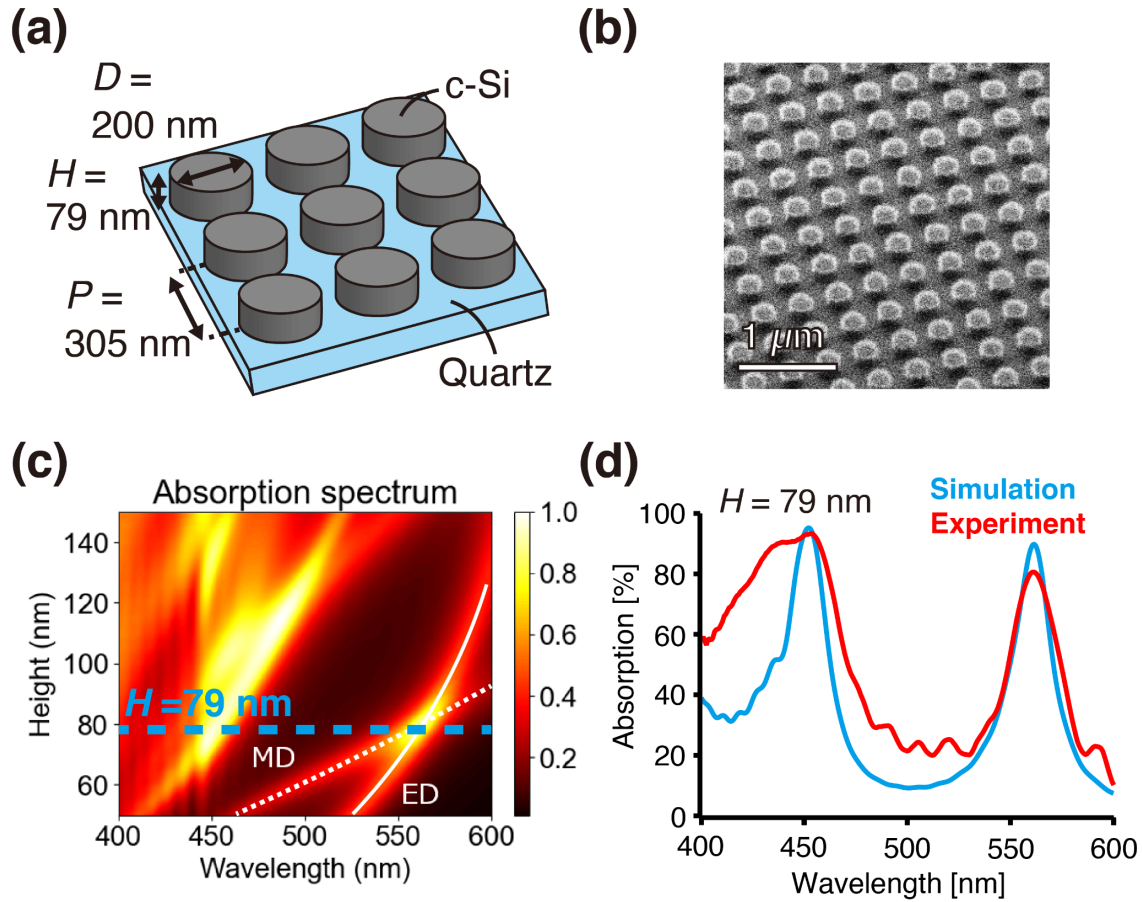

Figure 2: Absorption properties of the silicon metasurface. (a) Structure design schematic. (Abbreviation, D: diameter, H: height, P: period, c-Si: crystalline-silicon) (b) SIM image of the fabricated silicon metasurface. (c) Absorption spectral map of silicon metasurface for different heights of the meta-atoms. The diameter and the period of silicon nanodisc meta-atoms are 200 nm and 305 nm, respectively. The surrounding medium is air ($n$=1). Blue dotted line indicates the height of 79 nm. White solid and dotted lines indicate the locations of ED and MD modes, respectively. (d) Absorption spectra agree well between simulation (Blue line) and experiment (Red line), showing a resonance peak of Huygens' dipole at 561 nm.

## 2.2 Calculation of photothermal nonlinear scattering

We theoretically investigated the photothermal nonlinear scatterings from the silicon metasurface under the illumination at the resonance peak wavelength. The first step is to confirm the scattering spectra shift at high temperature, and the second step is to link temperature rise with excitation intensity. We started from calculating the dependencies of forward and backward scattering spectra on the temperature of the silicon metasurface. In the FDTD simulation, the beam size is set to 4.4 $\mu$m so that 9×9 silicon meta-atoms are illuminated. In order to induce high absorption by the optical interactions between meta-atoms [24], at least 5×5 silicon nanodiscs meta-atoms must be included in the beam diameter, as shown in Supplementary Materials (Figure S3). Temperature dependencies of complex refractive index from crystalline silicon were previously measured by ellipsometry with a heating stage (M-2000, J.A. Woolam) [17].

Figure 3a shows the calculated forward scattering spectrum of silicon metasurface at the temperature of 300 K, 500 K, and 700 K. When the temperature of the silicon metasurface is 300 K, the forward scattering spectrum has a peak at 561 nm, indicating the dominance of forward scattering, as the signature of the Huygens' metasurface. However, at the temperature of 700 K, the peak wavelength shows a red-shift of ~12 nm because of the thermo-optical refractive index variation of crystalline silicon, and thus the scattering efficiency at the wavelength of 561 nm decreases to nearly half of that at 300 K. It should be noted that the peak intensity value of the scattering spectra decreases as temperature increases, probably because thermal energy dissipation grows due to the enlarged imaginary refractive index. Figure 3b shows the backward scattering spectra at the corresponding temperatures. At 300K, complementary to the forward scattering spectrum in Figure 3a, the backward scattering is efficiently suppressed at 561 nm in the Huygens' metasurface. As the temperature increases to 700 K, the backward scattering spectrum shows ~9 nm red-shift, and the backward scattering efficiency at 561 nm increases to about 2.5-fold of that at 300 K.

Figure 3c shows the dependencies of forward scattering intensity and temperature of silicon metasurface, versus the excitation intensity at 561 nm, which is the resonance wavelength of the metasurface. The temperature and the scattering intensity were calculated by the FDTD method with an iterative calculation, which discretized the temperature rising process and the photothermally induced variation of absorption cross-section of silicon metasurface. The heat flux of the laser illumination was converted into temperature by Fourier's heat equation, and we sequentially calculated absorption and scattering cross-sections at each temperature, until the temperature reached steady-state. Detailed simulation algorithms and the thermal parameters are reported in our previous paper [17].

When the excitation intensity is lower than ~0.2 mW/$\mu$m$^2$, forward scattering intensity linearly increases. However, when the excitation intensity is over the threshold and the temperature reaches ~340 K, the forward scattering intensity starts to show a sub-linear trend due to the red-shift of the scattering spectrum, as shown in Figure 3a. On the other hand, in Figure 3d, the backward scattering intensity keeps almost zero, until the excitation intensity grows to ~0.6 mW/$\mu$m$^2$, where the temperature reaches ~440 K. From Figure 3b, between 300K and 500K, the backward scattering spectrum exhibits both red-shift and dip value reduction. The reason that backward scattering at 561 nm does not increase with excitation intensity is that the dip value reduction dominates before reaching 500K. When the excitation intensity is larger than ~0.6 mW/$\mu$m$^2$, the backward scattering starts to grow with the excitation intensity because the increase of backward scattering by the spectral shift dominates when the temperature exceeds ~500 K. These simulation results confirm the feasibility of all-optical photothermal control of scattering directionality. It is also interesting to note that the forward and backward scattering intensities respectively show gradual reverse-saturation and saturation effects, when the excitation intensity rises up to ~2.0 mW/$\mu$m$^2$ and the temperature reaches ~920 K, as shown in Supplementary Materials (Figure S4).

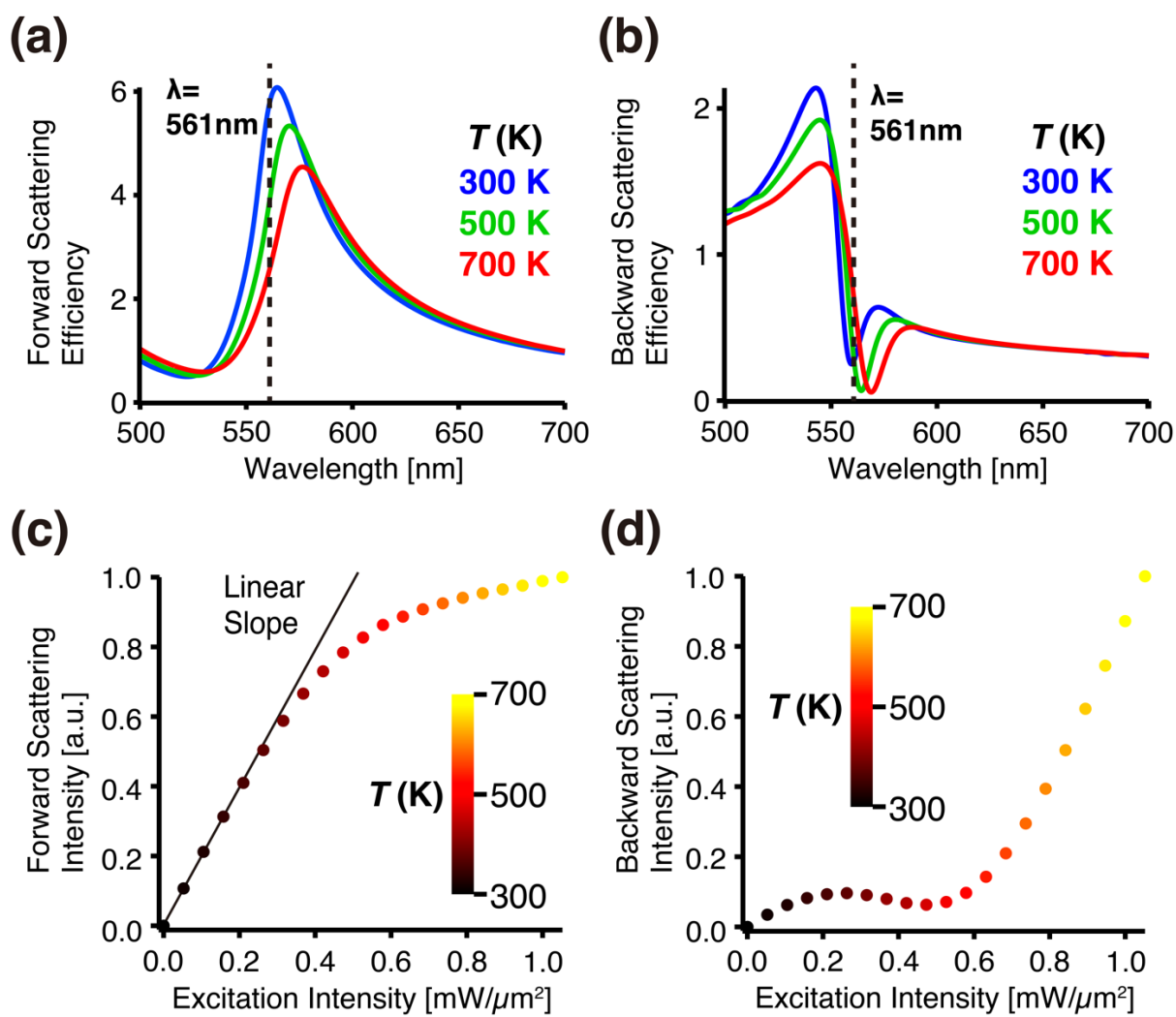

Figure 3: Theoretical calculations on the photothermal nonlinear scattering from the silicon metasurface. (a,b) Scatterings spectra of (a) forward and (b) backward scatterings at the temperature of 300 K (Blue), 500 K (Green), and 700 K (Red). The vertical axis indicates the scattering efficiency per a single silicon meta-atom, which is calculated by the ratio of scattering cross-section and geometric cross-section of the single meta-atom. (c,d) Dependences of (c) forward and (d) backward scattering intensities on the excitation intensity, at the excitation wavelength of 561 nm. The color of the plot indicates the equilibrium temperature (K) under light irradiation. The solid line in (c) indicates a linear slope.

**2.3 Experimental demonstration**

Figure 4 reports the experimental observation of the scattering spectra and photothermal nonlinear scattering dependence from the silicon metasurface. Figure 4a is the schematic of the optical setup in this experiment. The forward scattering spectrum was obtained by a dark-field illumination spectroscopy system. A halogen lamp (LG-PS2, Olympus) illuminated the sample through a dark-field condenser lens with an NA of 0.8-0.92 (U-DCD, Olympus), and the forward scattering from the sample was collected by a dry objective lens with an NA of 0.4 (UPlanSApo, Olympus). The scattering light was relayed to a spectrometer (Kymera 328i, Andor). On the other hand, the backward scattering spectrum was acquired by a supercontinuum laser (SC-450-PP, Fianium). The illumination intensity is adjusted by a tunable ND filter at the laser exit. The supercontinuum beam was focused on the sample by the dry objective lens. The position of the illumination spot was controlled by two-axis galvanometer mirrors. The backward scattering from the sample was collected by the same objective lens and was detected by the same spectrometer. During this measurement, the continuous-wave (CW) laser light source oscillating at 561 nm (Cobolt Jive, HUBNER Photonics) was turned off.

Figures 4b and 4c respectively show forward and backward scattering spectra of the silicon metasurface, measured at room temperature. Both spectra are normalized with the spectra of individual light sources. While the forward scattering spectrum has a peak at 561 nm, the backward scattering shows a dip at the same wavelength, in good agreement with the simulation results in Figures 3a and 3b.

After experimentally confirming the scattering spectra, we measured the relationship between excitation intensity and scattering intensities from silicon metasurface with a single wavelength illumination of 561 nm, in order to observe photothermal nonlinearity of forward and backward scattering intensities. As shown in Figure 4a, the excitation light source was a CW laser oscillating at 561 nm. The excitation intensity was controlled through the combination of a half-wave plate and a polarizing beam splitter. The excitation beam was focused on the sample by the NA 0.4 dry objective lens. The forward scattering light was collected by a dark-field condenser lens and detected by a photomultiplier tube (PMT) in the transmission path. The backward scattering light was collected by the same objective lens and detected by another PMT in the reflection path after passing through a confocal pinhole. The beam size of the Gaussian excitation focal spot was 4.4 $\mu$m, which corresponds to the beam diameter in the simulation. The scattering signal was acquired by continuously scanning the excitation focus spot on the silicon metasurface with the dwell time of 12 $\mu$s at each point, to avoid excess heat accumulation, and also characterize the wide area of nonlinear responses on the silicon metasurface as shown in Figure S5 of Supplementary Materials. Note that the heating/cooling effect of silicon nanostructure requires only the order of nanosecond scale, according to our previous research [17], and this heating/cooling time is much faster than the dwell time of the focus spot (12 $\mu$s). Therefore, we measured the scattering intensity after the nanostructure reached thermal equilibrium at each scanning position.

Figure 4d is the experimentally measured relationship between forward scattering and excitation intensities at the excitation wavelength of 561 nm, which corresponds to the resonance peak of our perfect absorber. At excitation intensity below ~0.2 mW/$\mu$m$^2$, the forward scattering intensity follows a linear slope. As the excitation intensity increases to more than ~0.2 mW/$\mu$m$^2$, the scattering intensity starts to deviate from the linear slope and shows sub-linear dependence on excitation intensity. At the excitation intensity of ~0.8 mW/$\mu$m$^2$, the deviation ratio from the linear trend reaches -40%, which is calculated by the ratio between the deviation of measured scattering intensity from the extrapolated linear slope and the expected scattering intensity from the linear slope [17]. Figure 4e is the graph of backward scattering intensity. The scattering intensity shows a proportional increase to excitation intensity at the excitation intensity lower than 0.12 mW/$\mu$m$^2$, different from that in Figure 3d. At such a low excitation intensity (i.e. room temperature), our fabricated silicon metasurface exhibits nearly zero backward scattering at the wavelength of 561 nm, as shown in Figure 4c, and this quite weak scattering signal might be easily covered by the light scattering background from the substrate, which is inevitable in backward scattering measurement. Therefore, the behavior of backward scattering intensity at low excitation regions, especially the effect of scattering cross-section reduction (Figure 3b), was not clearly visualized in the experiment, unlike the calculation result in Figure 3d. The scattering intensity starts to show steep increases in excitation intensity, when the excitation intensity exceeds 0.12 mW/$\mu$m$^2$. In addition, by increasing the excitation intensity by more than ~0.6 mW/$\mu$m$^2$, the backward scattering intensity starts to show saturation. The deviation ratio reaches +115 % at the excitation intensity of ~0.8 mW/$\mu$m$^2$. These experimental results confirmed our proposed model of photothermally modulated scattering responses from Huygens' metasurface, agreeing reasonably well with the overall trend of graphs in the simulation (Figures 3c and 3d).

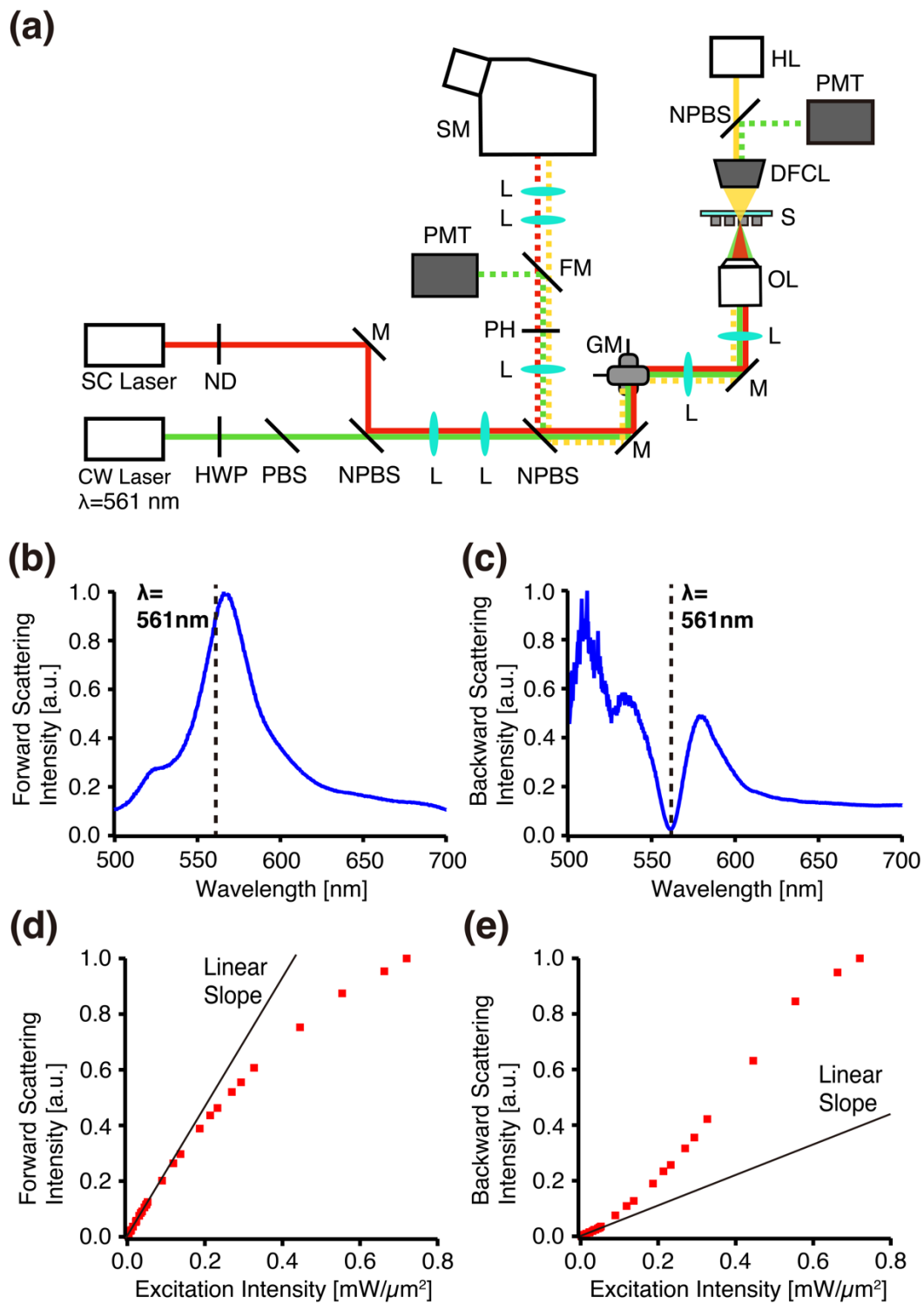

Figure 4: Experimental measurement of photothermal nonlinear scattering from silicon metasurface. (a) Schematic of the optical setup. (Abbreviation, CW: continuous-wave, SC: Super-continuum, ND: neutral density, HWP: half-wave plate, PBS: polarization beam splitter, NPBS: non-polarization beam splitter, M: mirror, L: lens, GM: galvanometer mirror, OL: objective lens, S: sample, DFCL: dark-field condenser lens, HL: halogen lamp, PMT: photomultiplier tube, PH: pinhole, FM: flippable mirror, SM: spectrometer). (b,c) Scattering spectra of (b) forward and (c) backward scatterings, measured at room temperature. (d,e) Dependences of (d) forward and (e) backward scattering intensities on the excitation intensity. The solid line indicates a linear slope. The excitation wavelength is 561 nm.

## 3. Discussion and conclusion

Compared to our previous results of single silicon nanoblocks, the high absorption of metasurface significantly reduces the excitation intensity threshold to induce photothermal nonlinear scattering by an order of magnitude [17]. To further improve the efficiency of the photothermal scattering nonlinearity in silicon metasurface, more efficient light-heat conversion via the increase of absorption in dielectric metasurface is highly desirable. In principle, silicon metasurface has the potential to achieve 100% absorption, i.e., metamaterial perfect absorber (MPA), in the condition of degenerate critical coupling [22,25]. However, in our metasurface, the condition of critical coupling is not perfectly fulfilled because of the insufficient imaginary permittivity of crystalline silicon at the wavelength of 561 nm. One approach to solve this problem is to add amorphous silicon caps on the silicon disc meta-atoms to increase the material loss, although the fabrication process becomes complex [20].

A similar concept of optical modulation of metal-dielectric-metal (MDM)-based MPA[26–28] using a photothermal variation of scattering spectra has been demonstrated [29]. However, because the MDM-based MPA is opaque due to the metallic coating, its direction of optical control is limited to only backward scattering. The maximal nonlinear deviation ratio is ~-30% at the average power of ~5.5 mW [29]. On the other hand, all-dielectric MPA allows us to control both forward and backward scattering, which are more flexible for future device applications. In our results, we experimentally achieved much larger nonlinear deviation ratios in not only the backward (+115%), but also the forward (-40%) scattering at a similar excitation intensity. Furthermore, the all-dielectric resonator potentially provides a better Q-factor at the resonance wavelength because of its low loss nature, and it is possible to further enhance nonlinearity through spectral red-shift via photothermal interaction, as demonstrated in our research. Last but not least, compared to MDM-based MPA, the high melting point and the CMOS fabrication compatibility of silicon are important advantages for mass device production.

In this paper, we have demonstrated the photothermal nonlinearity of forward and backward scattering intensities from a silicon quasi-perfect absorbing metasurface under the excitation at the resonance wavelength of the metasurface. We designed and fabricated a crystallized silicon metasurface that mostly fulfills the condition of degenerate critical coupling, exhibiting both properties of scattering directionality via Huygens' dipoles and high absorption at >80%. We theoretically found that both peak wavelengths of forward and backward scattering spectra from the silicon metasurface showed red-shift with heating. The calculation results also showed that forward and backward scattering cross-sections respectively exhibit a decrease and increase to the excitation intensity through the photothermal effect under resonant excitation. We experimentally confirmed the photothermal nonlinear control of Huygens' dipoles by observing forward and backward scattering intensities, agreeing well with the calculation results. Our research opens up the approach of all-optical dynamic control of scattering direction based on the photothermal effect of the all-dielectric metasurface. This research also demonstrated the feasibility of an active all-optical switching device by using notably thin dielectric film with the sub-wavelength thickness of only λ/7, where λ is the operation wavelength, which would significantly contribute for the miniaturization of the optical devices to induce light modulation, and provide new idea in the field of silicon photonics.

ASSOCIATED CONTENT

**Supplementary Materials**

Additional data and figures including, the relationship between the period of silicon nanodisc meta-atoms and absorption, optical setup for spectral measurement, dependence of the number of illuminated silicon discs meta-atoms on the absorption, theoretically calculated dependences of forward and backward scattering intensities from silicon metasurface on the excitation intensity for a high excitation intensity region, forward and backward scatterings images of silicon metasurfaces at various excitation intensities.

AUTHOR INFORMATION

**Corresponding Authors**

- Junichi Takahara (takahara@ap.eng.osaka-u.ac.jp)

Graduate School of Engineering and Photonics Center, Osaka University, 2-1 Yamadaoka, Suita, Osaka 565-0871, Japan

- Shi-Wei Chu (swchu@phys.ntu.edu.tw)

Department of Physics and Molecular Imaging Center, National Taiwan University, 1, Sec 4, Roosevelt Rd., Taipei 10617, Taiwan; Brain Research Center, National Tsing Hua University, 101, Sec 2, Guangfu Road, Hsinchu 30013, Taiwan

**Author Contributions**

The manuscript was written through the contributions of all authors. All authors have given approval to the final version of the manuscript.

**Funding Sources**

Ministry of Science and Technology, Taiwan (MOST 109-2112-M-002-026-MY3, 111-2321-B-002-016, 111-2119-M-002-012-MBK); as well as supported by The Featured Areas Research Center Program (NTHU) and NTU Higher Education Sprout Project (NTU-110L8809) within the framework of the Higher Education Sprout Project co-funded by the MOST and the Ministry of Education, Taiwan (MOE).; This work was supported by Japan Society for the Promotion of Science (JSPS) Core-to-Core Program and KAKENHI JP19H02630. We would like to thank Shin-Etsu Chemical Co., Ltd. for donating silicon-on-quartz substrates.